\documentclass[10pt]{article}
\usepackage[dvips]{graphicx}
\usepackage{amssymb}
\usepackage{amsmath}
\usepackage{amsfonts}
\newtheorem{dfn}{Definition}
\newtheorem{prop}{Proposition}
\newtheorem{theorem}{Theorem}
\newtheorem{remark}{Remark}
\newtheorem{inference}{Inference}
\usepackage{mathabx}
\def\one{1\hskip -1mm{\rm l}}
\begin{document}

\begin{center}
{\Large \bf \sf
      Bi-orthogonal eigen-spinors related to $\mathcal{T}$-pseudo-Hermitian Pauli Hamiltonian : Time reversal and Clifford Algebra}

\bigskip\vspace{1cm}
{\sf Arindam Chakraborty\footnote{e-mail: arindam.chakraborty@heritageit.edu}}

\bigskip

{\em $^1$Department of Physics, Heritage Institute of Technology, Kolkata-700107, India}

\end{center}

\vskip 20pt
\begin{center}
{\bf Abstract}
\end{center}

\par A set of two-parameter bi-orthogonal eigen-spinors has been constructed from a $\gamma$- deformed pseudo-Hermitian extension of Pauli Hamiltonian and its Hermitian conjugate. The Hamiltonians thus obtained are iso-spectral to the original Pauli Hamiltonian. A pair of spin-projection operators has been constructed as an essential ingredient of a possible bi-orthogonal quantum mechanics. An analogue of Kramers theorem in pseudo-Hermitian setting has also been inferred in a conjectural sense. The properties of time reversal and bi-orthogonality have been elaborated  in the frame work of Clifford algebra $(Cl_3(R))$, where the spinors have been viewed as elements of left ideal and the relevant inner-products are understood in terms of different involutions leading to elements of a division ring.  The whole process of present construction is found to be based on both direct and time reversed $Cl_3$ generators. A new variant of Kustaanheimo-Stiefel transformation has been introduced with the help of spinor operator.

AMS subject classification(2010):81R25

 {\bf  Keywords:} Bi-orthogonal Spinor, pseudo-Hermitian Hamiltonian, time reversal, Kramers theorem,  involution, Spinor operator, Kustaanheimo-Stiefel.

\section{Introduction}
Recently, bi-orthogonal quantum mechanics has become a major area of investigation in view of its close relation to pseudo-Hermitian Hamiltonians with real spectrum. It has been shown by Brody \cite{brody14} that a parallel formulation of quantum mechanics is possible at least in finite dimensional Hilbert space taking a complete set of bi-orthogonal states into account. The early attribution of space-time reflection symmetry to the reality of the eigenvalue by Bender \cite{bender98, bender99, bender02}, and many others \cite{mosta02, brody16, znojil04} has later been firmly grounded by Mostafazadeh \cite{mosta02a, mosta02b} with the introduction of more general notion of pseudo-Hermiticity. \textbf{A Hamiltonian $H$ is said to be $\eta$-pseudo-Hermitian if there exists at least one such invertible operator $\eta$ so that $\eta^{-1}H\eta=H^{\dagger}$ where, $H^{\dagger}$ is the Hermitian conjugate of $H$ in a given sense of inner-product.} Very often,  it is the space-time reflection symmetry or so called $\mathcal{PT}$-symmetry that provides the necessary physical ground behind any relevant choice of $\eta$. Pseudo-Hermitian systems have been studied in different contexts ranging from the systems with complex potentials,  resonance phenomena associated to nuclear, atomic or molecular systems, nano-structured materials or condensates, to even the systems which are not so quantum mechanical in sense but their physical behaviour is quite amenable to quantum language (for example classical statistical mechanical systems, biological systems with diffusion, light propagation in wave guides) and many other fields where even the conventional quantum mechanics has already shown success \cite{moise11, baga15, bender20}. On the other hand, apart from the development of formal methods \cite{brody14, curt07} bi-orthogonal system has got special attention in the study of super-symmetry, Lie super-algebra and quantum mechanics over Galois field etc. \cite{curt07a, houn17, oscar18, chang13}

The present report deals with a non-Hermitian version of so called Pauli Hamiltonian that describes the behaviour of a single electron in a magnetic field $\vec{B}$. The usual expression of interaction Hamiltonian $H_{\rm{int}}=\vec{\sigma}\cdot\vec{B}$ \cite{Hl99, Da08} has a $2\times 2$ matrix representation which is Hermitian in the conventional sense of inner-product $\langle u\vert v\rangle=u_1^{\star}v_1+u_2^{\star}v_2$ for any two vectors $\vert u\rangle, \vert v\rangle\in\mathbf{C}^{2}$. Here, $\vec{\sigma}=\{\sigma_1, \sigma_2, \sigma_3\}$ represents Pauli spin matrices. \textbf{In the language of real Clifford algebra $Cl_3(R)$ \cite{Lo01, vaz16} the Hamiltonian $H_{\rm{int}}$ is nothing but the expression of the magnetic field in terms of the generators $\{e_1, e_2, e_3\}\equiv \{\sigma_1, \sigma_2, \sigma_3\}$ acting as matrix valued $1$-blades}. Similar Hamiltonians involving spin-orbit coupling has been discussed by Chakraborty et. al. \cite{chakraborty22} with the introduction of generalized Clifford momentum. Clifford algebra in particular and geometric algebra in general have been widely used in formulating various physical theories \cite{doran03} like classical mechanics \cite{hestenes02}, electrodynamics \cite{jance89}, relativity \cite{hestenes15} and even in current areas of interest like quantum information theory \cite{cafaro11}.

Unlike its Hermitian counterpart our present Hamiltonian is written in terms of a new set of generators obtained from a set of bi-orthogonal vectors and two of them are non-Hermitian in conventional sense. This makes the interaction Hamiltonian non-Hermitian still admitting real eigenvalues. It is found that the same Hamiltonian is $\mathcal{T}$-pseudo-Hermitian under the action of a suitably chosen fermionic time reversal operator $\mathcal{T}$ \cite{geru18}.

In order to formulate a possible bi-orthogonal quantum system we consider a special case of the interaction Hamiltonian when $B_3=0$. \textbf{The eigen-spinors of the Hamiltonian and its Hermitian conjugate constitute a two-parameter-bi-orthogonal system in various senses and they have been studied as left ideals of the related Clifford algebra $Cl_3$. The ideal has been constructed with the help of different types of involution available in the framework of $Cl_3$ and what may be called time-reversed generators associated to it}.

The fermionic time-reversal operator acts as an anti-involution i. e.; $\mathcal{T}^2=-\mathbf{1}$ and it has been extensively discussed in \cite{geru18, jones10, bender11, jones14, choutri14}. One of the important features of such time reversal operator is the \textbf{Kramers degeneracy theorem} \cite{klein51, scharf87, kruthoff19, rosch83, konstantin20, revaz09, chen22, lieu22} which stems from the existence of simultaneous eigenstates of the Hamiltonian and the time reversal operator. In the present scenario we observe that the time reversal operator does not commute with the pseudo-Hermitian Hamiltonian thus prohibiting the possibility of Kramers' degeneracy.  However, a modified version of the said theorem has been obtained at least for the present case. The time reversal operator is of particular interest in the present occasion especially in the context of its close association with various \textbf{involutions} in Clifford algebra and inner-product associated with specific automorphism group. The importance of the time reversal operator $\mathcal{T}$ has been emphasized by recognizing spinor as the left ideal of the $Cl_3$ algebra and the relevant inner-product has been represented in the frame-work of a division ring. Finally,  a kind of spinor operator  has been constructed. Such operator has application in various physical \cite{cahill90, chen00, helff96, ino93} and mathematical contexts \cite{vaz19, El07}.

The article is organised in the following way : Section-\ref{auerbach}  discusses the construction of $\gamma$-deformed Clifford generators starting from a one-parameter bi-orthogonal system of vectors. A linear transformation is introduced to construct such a set starting from a orthogonal set. A set of two-parameter bi-orthogonal eigen-spinors
has been constructed in section-\ref{bi}. A couple of projection operators has also been introduced as a necessary ingredient for a possible bi-orthogonal quantum mechanics as envisaged by Brody \cite{brody14}.  In section-\ref{pseudo} the connection between fermionic time reversal and various involutions in Clifford algebra has been discussed and their correspondences with bi-orthogonality have been shown. This section concludes with the construction of \textbf{spinor operator} leading to a new variant of \textbf{Kustaanheimo-Stiefel transformation}\cite{cahill90, chen00} .

\section{Auerbach Bi-orthogonal System and  $Cl_{3}$ generators}\label{auerbach}

Starting from two orthogonal vectors $\{\vert u_j\rangle =
\frac{1}{\sqrt{2}}\left(\begin{array}{c}
	1 \\
	(-1)^{j-1}
\end{array} \right): j = 1,2\}\in \mathbf{C^2}$, it is possible to write Pauli matrices $\{\sigma_m : m=1, 2, 3\}$ in the following form
\begin{equation}\label{ortho1}
	\sigma_m=\frac{i^{m+1}}{2}\sum_{j, k=1}^{2}c_{jk}^{(m)}\vert u_j\rangle\langle u_k\vert,
\end{equation}
where, $i=\sqrt{-1}$, $c_{jk}^{(1)} = (-1)^j \delta_{jk}$ and $c_{jk}^{(3)} = (-1)^j c_{jk}^{(2)} = (1 -(-1)^j c_{jk}^{(1)})$. These matrices are Hermitian in the sense of conventional inner-product $\langle v_j\vert v_k\rangle=(c_j^{(1)})^{\star}c_k^{(1)}+(c_j^{(2)})^{\star}c_k^{(2)}$ defined on $\mathbf{C^2}$.

\begin{dfn}
	The Clifford algebra $Cl_{3}(\mathbf{R})$ is a real associative algebra equipped with the operation $ab=a\cdot b+a\wedge b$ and admitting the set of generators $\mathcal{G}^0=\{1; e_1, e_2, e_3; e_{12}, e_{23}, e_{31}; e_{123}\}$ satisfying the following relations:
	\begin{eqnarray}
		e_i^2&=&1\:\:\forall\:\: i=1,2, 3.\nonumber\\
		e_ie_j&=&-e_je_i\:\:\forall\:\: i\neq j
\end{eqnarray}
\end{dfn}

Identifying $\{\sigma_m : m=1, 2, 3\}=\{e_m : m=1, 2, 3\}$, The generators for $Cl_{3}(\mathbf{R})$ can be written as $\mathcal{G}^0=\{\mathbf{1}_2; \sigma_1, \sigma_2, \sigma_3, i\sigma_3; i\sigma_1, i\sigma_2; i\mathbf{1}_2\}$ and the isomorphism $Cl_3\cong R\bigoplus R^3\bigoplus\bigwedge^2R^3\bigoplus\bigwedge^3R^3$ is possible with the exterior algebra $\bigwedge R^3$ \cite{Lo01}.

The method mentioned in equation-\ref{ortho1} can be extended for a biorthogonal system with the following definition

\begin{dfn}
	Two pairs of vectors $\{\vert\phi_j\rangle : j=1,2\}$ and $\{\vert\chi_j\rangle : j=1,2\}\in\mathbf{C^2}$ are said to be bi-orthogonal if $\langle\phi_j\vert\chi_k\rangle=\delta_{jk}$.
\end{dfn}

The condition of bi-orthoganality in $C^2$ can be given by the following theorem

\begin{theorem}
	Given any pair of vectors
	$\{\vert v_j\rangle =
	\left(\begin{array}{c}
		c_j^{(1)} \\
		c_j^{(2)}
	\end{array} \right): j = 1,2\}\in \mathbf{C^2}$ two pairs of vectors $\vert\phi_j\rangle=T\vert v_j\rangle$ and $\vert\chi_j\rangle=(T^{-1})^{\dagger}\vert v_j\rangle$ constitute a bi-orthogonal system under the action of a transformation $T = (\cos \frac{\theta}{2} ){\bf 1}_2 +
	2(\cos \frac{\phi}{2}  \sin \frac{\theta}{2} )\sigma_1 - 2(\sin \frac{\phi}{2}  \sin \frac{\theta}{2} )\sigma_2$  provided $\langle v_j\vert v_k\rangle=(c_j^{(1)})^{\star}c_k^{(1)}+(c_j^{(2)})^{\star}c_k^{(2)}=0\:\:\forall\:\: j\neq k$.
\end{theorem}

\textbf{Proof} :
	Since $T=T^{\dagger}$ with the present sense of inner-product $\langle\phi_j\vert\chi_k\rangle=\langle v_j\vert T(T^{-1})^{\dagger}\vert v_k\rangle=\langle v_j\vert T(T^{\dagger})^{-1}\vert v_k\rangle=\langle v_j\vert v_k\rangle$. Hence the theorem follows from the notion of bi-orthogonal system stated above. $\square$.

\begin{remark}
Acoording to Auerbach such a bi-orthogonal system is always available in any finite dimensional Banach  space \cite{hajek08}
\end{remark}

Now, taking $\vert v_j\rangle=\vert u_j\rangle$, $\phi=\pi$ and $\gamma=\sqrt{1-\omega^2}=\sin\theta$ we get
\begin{equation}
	\sigma_m^{\gamma} = \frac{i^{m+1}}{2}\sum_{j, k=1}^2{c_{jk}^{(m)}}\vert\phi_j\rangle \langle\chi_k \vert : m = 1, 2, 3
\end{equation}

Writting explicitly

\begin{equation}\label{gamma}
	\sigma_1^{\gamma}=\omega^{-1}\left(\begin{array}{cc}
		-i\gamma & 1 \\
		1 & i\gamma
	\end{array} \right)=e_1^\gamma, \:\:\:\sigma_2^{\gamma}=\left(\begin{array}{cc}
	0 & -i \\
	i & 0
\end{array} \right)=e_2^\gamma \:\:{\rm{and}} \:\:\:\sigma_3^{\gamma}=\omega^{-1}\left(\begin{array}{cc}
1 & i\gamma \\
i\gamma & -1
\end{array} \right)=e_3^\gamma
\end{equation}

It is obvious to note that $\{\sigma_1^{\gamma}, \sigma_3^{\gamma}\}$ are non-hermitian in the conventional sense of inner-product.In the spirit of Definition-1 a new set of $Cl_3$ generators $\mathcal{G}^{\gamma}=\{1, e_1^{\gamma}, e_2^{\gamma}, e_3^{\gamma}, e_{12}^{\gamma}, e_{23}^{\gamma}, e_{31}^{\gamma}, e_{123}^{\gamma} \}$ can be formed {\footnote{Possibility of a similar set of generators has been hinted in \cite{Lo01}}}. Such generators resemble the previous one if $\gamma=0$. The algebra $Cl_3$ has the decomposition $Cl_3=Cl_3^+\bigoplus Cl_3^-$
 called even and odd parts respectively. $Cl_3^+$ is non-commutative and isomorphic to quaternion algebra $\mathbf{H}$. The generators of $Cl_3^+$ is given by $\mathcal{G}_+^{\gamma}=\{1, e_{12}^{\gamma}, e_{23}^{\gamma}, e_{31}^{\gamma}\}$.

\section{Spectrum of the Hamiltonian and two-parameter bi-orthogonal eigenspinors}\label{bi}
In view of the equation-\ref{gamma} let us write the interaction Hamiltonian $H_\gamma$ with $B_3=0$  and its corresponding Hermitian conjugate $H_\gamma^\dagger$ in terms of the new generators
\begin{equation}\label{hamil}
H_{\gamma}=e_1^{\gamma}B_1+e_2^{\gamma}B_2\:\:{\rm{and}}\:\: H_\gamma^\dagger=e_1^{-\gamma}B_1+e_2^{-\gamma}B_2=H_{-\gamma}.
\end{equation}

In order to validate the possibility of a bi-orthogonal quantum mechanics relating to a non-Hermitian Hamiltonian we make use of the following results.

\begin{theorem}\label{biortho1}
Let, $\mathcal{H}$ be a finite dimensional Hilbert space and $H$ be a non-Hermitian Hamiltonian acting on $\mathcal{H}$ with the decomposition $H=H_0+i\Gamma$, where, $H_0=H_0^{\dagger}$ and $\Gamma=\Gamma^{\dagger}$.  If  $\{\vert u_n\rangle\}$ be the eigenstates corresponding to all non-degenerate eigenvalues $\{\lambda_n\}$ of $H$ and $\{\vert v_n\rangle\}$ be the eigenstates corresponding to all non-degenerate eigenvalues $\{\mu_n\}$ of $H^\dagger$ the following results hold

\begin{eqnarray}
\langle u_m\vert u_n\rangle &=&\frac{-2i}{\lambda_m^\star-\lambda_n}\langle u_m\vert \Gamma\vert u_n\rangle=\frac{2}{\lambda_m^\star+\lambda_n}\langle u_m\vert H_0\vert u_n\rangle,\nonumber\\
\langle v_m\vert v_n\rangle &=&\frac{-2i}{\mu_m^\star-\mu_n}\langle v_m\vert \Gamma\vert v_n\rangle=\frac{2}{\mu_m^\star+\mu_n}\langle v_m\vert H_0\vert v_n\rangle\:\:{\rm{and}}\nonumber\\
\langle u_m\vert v_n\rangle &=& 0\:\:\forall\:\:m\neq n.
\end{eqnarray}
\end{theorem}

Proof : We can write two eigenvalue equations
\begin{equation}
H\vert u_n\rangle=(H_0+i\Gamma)\vert u_n\rangle=\lambda_n\vert u_n\rangle\:\:{\rm{and}}\:\:\langle u_n\vert H^{\dagger}=\lambda_n^\star\langle u_n\vert
\end{equation}
On similar ground the following equations are also possible
\begin{equation}
H^{\dagger}\vert v_n\rangle=(H_0-i\Gamma)\vert v_n\rangle=\mu_n\vert v_n\rangle\:\:{\rm{and}}\:\:\langle v_n\vert H=\mu_n^\star\langle v_n\vert
\end{equation}
In view of the above two results we can write
\begin{eqnarray}
\langle u_m\vert H\vert u_n\rangle=\langle u_m\vert (H_0+i\Gamma)\vert u_n\rangle=\lambda_n\langle u_m\vert u_n\rangle\nonumber\\
\langle u_m\vert H^\dagger\vert u_n\rangle=\langle u_m\vert (H_0-i\Gamma)\vert u_n\rangle=\lambda_m^\star\langle u_m\vert u_n\rangle
\end{eqnarray}
 Adding the above two equations we get
 \begin{equation}\label{crucial}
 (\lambda_n+\lambda_m^\star)\langle u_m\vert u_n\rangle=2\langle u_m\vert H_0\vert u_n\rangle
 \end{equation}
Subtracting the first from the second we get
 \begin{equation}\label{crucial1}
 (\lambda^\star_m-\lambda_n)\langle u_m\vert u_n\rangle=-2i\langle u_m\vert \Gamma\vert u_n\rangle
 \end{equation}

Similar calculation may be repeated for $\vert v_n\rangle$ etc.
Finally, $\langle u_m\vert H^\dagger\vert v_n\rangle=\lambda_m^\star\langle u_m\vert v_n\rangle=\mu_n\langle u_m\vert v_n\rangle$. Since eigenvalues are non-degenerate $\langle u_m\vert v_n\rangle=0$ for all $m\neq n$. $\square$

\begin{remark}
In finite dimension the completeness of the bases $\{\vert u_n\rangle\}$ implies that of $\{\vert v_n\rangle\}$ . However, in infinite dimension this is no longer the case \cite{brody14}. Furthermore, equation-\ref{crucial} fails to work in tandem with the theorem-\ref{biortho1} if $H_\gamma$ and $H_\gamma^\dagger$ are iso-spectral with the $m$-th and $n$-th eigenvalues having same magnitudes but opposite signs, a situation which resembles precisely the case we are dealing with.  Similar trouble does not arise with equation-\ref{crucial1}.
\end{remark}

 The above two Hamiltonians in equation-\ref{hamil} are isospectral and have the eigenvalues $\lambda_\pm^\gamma=E_{\pm}=\pm\vert \vec{B}\vert=\pm B=\pm\sqrt{B_1^2+B_2^2}$.  Let, the corresponding eigenvalue equations for $H_{\pm\gamma}$ be
 \begin{eqnarray}\label{hamilmag}
 {\rm{(a)}}\:\:H_{\gamma}\vert\psi_\pm^{\gamma}\rangle=\pm B\vert\psi_\pm^{\gamma}\rangle\:\:{\rm{and}}\:\:{\rm{(b)}}\:\:H_{-\gamma}\vert\psi_\mp^{-\gamma}\rangle=\mp B\vert\psi_\mp^{-\gamma}\rangle
 \end{eqnarray}

 Taking adjoint on both sides equation-\ref{hamilmag}-(b) we get
\begin{eqnarray}\label{hamilmag1}
\langle \psi_\mp^{-\gamma}\vert H_{\gamma}=\mp B\langle\psi_\mp^{-\gamma}\vert
\end{eqnarray}

 Left multiplying equation-\ref{hamilmag}-(a) by $\langle \psi_\mp^{-\gamma}\vert$ and right multiplying equation-\ref{hamilmag1} by $\vert\psi_\pm^{\gamma}\rangle$ we get
 \begin{eqnarray}\label{hamilmag2}
 {\rm{(a)}}\:\:\langle \psi_\mp^{-\gamma}\vert H_\gamma\vert\psi_\pm^{\gamma}\rangle=\pm B\langle \psi_\mp^{-\gamma}\vert\psi_\pm^{\gamma}\rangle\:\:{\rm{and}}\:\:{\rm{(b)}}\:\:\langle \psi_\mp^{-\gamma}\vert H_\gamma\vert\psi_\pm^{\gamma}\rangle=\mp B\langle \psi_\mp^{-\gamma}\vert\psi_\pm^{\gamma}\rangle
 \end{eqnarray}

 Subtracting equation-\ref{hamilmag2}-(b) from equation-\ref{hamilmag2}-(a) we get $\pm 2B\langle \psi_\mp^{-\gamma}\vert\psi_\pm^{\gamma}\rangle=0$ resulting to $\langle\psi^{-\gamma}_-\vert\psi_+^{\gamma}\rangle=0=\langle\psi_+^{-\gamma}\vert \psi_-^{\gamma}\rangle$ thus justifying bi-orthogonality in the conventional sense of inner-product.

Explicit calculation shows
$\vert\psi^{\gamma}_+( \theta_+)\rangle =
\frac{1}{\sqrt{2}}\left(\begin{array}{c}
	e^{-i\theta_+} \\
	1
\end{array} \right)$ and $\vert\psi_-^{\gamma}( \theta_-) \rangle=
\frac{1}{\sqrt{2}}\left(\begin{array}{c}
-e^{i\theta_-} \\
1
\end{array} \right)$	
 are eigenvectors of $H_\gamma$ corresponding to the eigenvalues $\pm B$ and
$\vert\psi^{-\gamma}_+(\theta_-)\rangle =
	\frac{1}{\sqrt{2}}\left(\begin{array}{c}
		1 \\
		e^{-i\theta_-}
	\end{array} \right)$ and $\vert\psi_-^{-\gamma}( \theta_+)\rangle =
	\frac{1}{\sqrt{2}}\left(\begin{array}{c}
		-1 \\
		e^{i\theta_+}
	\end{array} \right)$ are eigenvalues of $H_{-\gamma}$.	
Here, $\theta_\pm(\gamma, \vec{B})=\tan^{-1}\frac{\pm\gamma B_1^2+ \omega^2BB_2}{ \omega BB_1\mp\gamma\omega B_1B_2}$.

\begin{remark}
	For $B_1=\pm B_2$, $\theta_\pm$ is independent of the magnetic field. \textbf{Furthermore the advange of choosing $\vert\gamma\vert< 1$ to obtain a set of bi-orthogonal pair to construct the generators of $Cl_3$ (see section-\ref{auerbach}) is reflected also in the expressions of bi-orthogonal states for $H_{\pm\gamma}$}. The expressions of $\theta_\pm$ represent \textbf{real angles} for such a choice. It is interesting to note that unlike Brody's consideration \cite{brody14} we have obtained a \textbf{two-paramter-bi-orthogonal set}  in terms of the parameters $\theta_{\pm}(\gamma, \vec{B})$. It is to be noted that $\theta_{\pm}(-\gamma, -\vec{B})=\theta_{\pm}(\gamma, \vec{B})$ and $\theta_\pm(\gamma, -\vec{B})+\theta_\mp(\gamma, \vec{B})=\theta_\pm(-\gamma, \vec{B})+\theta_\mp(\gamma, \vec{B})=0$.
\end{remark}

One can also construct the bi-orthogonal spinor projection operators
\begin{eqnarray}
	\Pi_+^{\gamma}=\frac{\vert\psi_+^{\gamma}\rangle\langle \psi_+^{-\gamma}\vert}{\langle \psi_+^{-\gamma}\vert\psi_+^{\gamma}\rangle}
	=\frac{1}{e^{-i\theta_+}+e^{i\theta_-}}\left(\begin{array}{cc}
	e^{-i\theta_+} &	e^{i(\theta_--\theta_+)} \\
	1 & e^{i\theta_-}	
	\end{array} \right)
\end{eqnarray}

and

\begin{eqnarray}
	\Pi_-^{\gamma}=\frac{\vert\psi_-^{\gamma}\rangle\langle \psi_-^{-\gamma}\vert}{\langle \psi_-^{-\gamma}\vert\psi_-^{\gamma}\rangle}
	=\frac{1}{e^{-i\theta_+}+e^{i\theta_-}}\left(\begin{array}{cc}
		e^{i\theta_-} &	-e^{i(\theta_--\theta_+)} \\
		-1 & e^{-i\theta_+}	
	\end{array} \right)
\end{eqnarray}

It is easy to verify that $\Pi_+^{\gamma}+\Pi_-^{\gamma}=\mathbf{1}_2$, $\Pi_+^{\gamma}\Pi_-^{\gamma}=0$, $(\Pi_\pm^\gamma)^2=\Pi_\pm^{\gamma}$ and $(\Pi_\pm^\gamma)^\dagger\neq\Pi_\pm^\gamma$. It is obvious to see that a similar bi-orthogonal formulation is also possible by choosing a set of projectors
$\Pi_+^{-\gamma}=\frac{\vert\psi_+^{-\gamma}\rangle\langle \psi_+^{\gamma}\vert}{\langle \psi_+^{\gamma}\vert\psi_+^{-\gamma}\rangle}$ and $\Pi_-^{-\gamma}=\frac{\vert\psi_-^{-\gamma}\rangle\langle \psi_-^{\gamma}\vert}{\langle \psi_-^{\gamma}\vert\psi_-^{-\gamma}\rangle}$. It is possible to write an equivalent spectral representation theorem (!) in the form $H_{\pm\gamma}=B\Pi_{+}^{\pm\gamma}-B\Pi_{-}^{\pm\gamma}$. We can thus prepare all the essential ingredients to construct a bi-orthogonal quantum mechanics as envisaged by Brody \cite{brody14}. However a Clifford algebraic interpretation of the set of projectors $\{\Pi_\pm^{\pm\gamma}\}$ can be given in terms of various involutions discussed in the next section (Remark-\ref{invo}).

\section{Pseudo-Hermiticity, fermionic time reversal and Clifford algebra}\label{pseudo}

The time reversal operator for $H_{\gamma}$ is given by $\mathcal{T}=e_{13}\mathcal{T}_0=-i\sigma_2\mathcal{T}_0$ whose action on spinor state $\vert\psi\rangle$ is given by
\begin{eqnarray}
	\mathcal{T}_0\vert\psi\rangle =\mathcal{T}_0\left(\begin{array}{c}
		\psi_1 \\
		\psi_2
	\end{array} \right)=\left(\begin{array}{c}
	\psi_1^{\star} \\
	\psi_2^{\star}
\end{array} \right),
\mathcal{T}\vert\psi\rangle=\mathcal{T}\left(\begin{array}{c}
	\psi_1 \\
	\psi_2
\end{array} \right)=\left(\begin{array}{c}
	-\psi_2^{\star} \\
	\psi_1^{\star}
\end{array} \right),\mathcal{T}^{-1}i\mathcal{T}=-i
\end{eqnarray}
and on $\{\sigma_j : j=1,2,3\}$ or $\{\sigma^\gamma_j :j=1,2,3\}$ are given by $\mathcal{T}^{-1}{\sigma_j}\mathcal{T}=-{\sigma_j}$ and $\mathcal{T}^{-1}{\sigma_j}^{\gamma}\mathcal{T}=-{\sigma_j}^{-\gamma}$ respectively. The operator $\mathcal{T}$ an anti-involution since $\mathcal{T}^2\vert\psi\rangle=-\vert\psi\rangle$.

It is easy to verify that under the action of $\mathcal{T}$, $H_{\gamma}$ is $\mathcal{T}$-pseudo-Hermitian that is
\begin{equation}
\mathcal{T}^{-1}H_{\gamma}\mathcal{T}=H_{\gamma}^{\dagger}\:\:\:{\rm{and}}\:\:{\rm{furthermore}}\:\:H_{\gamma}^{\dagger}=H_{-\gamma}.
\end{equation}

\begin{theorem}\label{schrodinger}
Given the time-dependent Schr\"odinger equation
\begin{eqnarray}
H\vert\psi(t)\rangle=i\partial_t\vert\psi(t)\rangle
\end{eqnarray}
satisfied by $\psi(t)\rangle$, $\mathcal{T}\vert\psi\rangle$ satisfies the time reversed Schr\"odinger equation with the Hamiltonian $H^\dagger=\mathcal{T}^{-1}H\mathcal{T}$.
\end{theorem}

\textbf{Proof}
Effecting time reversal $\mathcal{T}$ on both sides one can write
\begin{eqnarray}
\mathcal{T}H\mathcal{T}^{-1}\mathcal{T}\vert\psi(t)\rangle &=&\mathcal{T}i\mathcal{T}^{-1}\mathcal{T}\partial_t\mathcal{T}^{-1}\mathcal{T}\vert\psi(t)\rangle\nonumber\\
{\rm{or}}\:\:\mathcal{T}^{-1}H\mathcal{T}\vert\psi(t)\rangle &=&-i\partial_{-t}\mathcal{T}\vert\psi(t)\rangle\:\:{\rm{using}}\:\:\mathcal{T}^{-1}=-\mathcal{T}\nonumber\\
{\rm{or}}\:\:H^\dagger\vert\mathcal{T}\psi\rangle &=&i\partial_t\vert\mathcal{T}\psi\rangle\:\:\square
\end{eqnarray}

The following properties of $\mathcal{T}$ are very crucial

\begin{theorem}
$\mathcal{T}$ is anti-unitary, norm preserving and $\vert\psi\rangle$ and $\mathcal{T}\vert\psi\rangle$ are orthogonal.
\end{theorem}

\textbf{Proof}
Considering the obvious result
\begin{eqnarray}\label{timerev}
\langle\mathcal{T}\psi\vert\mathcal{T}\phi\rangle=\psi_1\phi_1^\star+\psi_2\phi_2^\star=\langle\phi\vert\psi\rangle.
\end{eqnarray}
 Putting $\vert\phi\rangle=\vert\psi\rangle$ in the equation-\ref{timerev}, $\langle\mathcal{T}\psi\vert\mathcal{T}\psi\rangle=\langle\psi\vert\psi\rangle$. This proves $\mathcal{T}$ is norm-preserving.

On the other hand, putting $\vert \phi\rangle=\mathcal{T}\vert\psi\rangle$ in equation-\ref{timerev} the l. h. s. gives $\langle\mathcal{T}\psi\vert\mathcal{T}^2\psi\rangle=-\langle\mathcal{T}\psi\vert\psi\rangle$ while the r. h. s.gives $\langle\phi\vert\psi\rangle=\langle\mathcal{T}\psi\vert\psi\rangle$ or $\langle\mathcal{T}\psi\vert\psi\rangle=0\:\:\square$.

 \vspace{.5cm}

It is to be mentioned that neither the operators $\{e_1^{\gamma}, e_2^{\gamma}\}$ nor the components of $\vec{B}$  have the said $\mathcal{T}$-pseudo-Hermiticity property at their individual level. However, the interaction Hamiltonian term adopts this property. \textbf{In Hermitian quantum mechanics Kramers' theorem tells that for fermionic systems with half-integer total spin where time reversal symmetry (TRS) is present, all energy levels are doubly degenerate.}{\footnote{However, Kramers did not connected this degeneracy of the energy levels with time-reversal symmetry \cite{geru18}}} Since, the time reversal symmetry implies the commutativity of $\mathcal{T}$ and the Hamiltonian and in the present case the operator $\mathcal{T}$ does not commute with the Hamiltonian $H_{\gamma}$ the conventional sense of Kramers degeneracy theorem cannot be expected. However, an alternative version of the same can be possible in the present context through the following observation.

It is obvious to note that both $H$ and $H^\dagger_\gamma$ are iso-spectral and their eigenvalues are $E_\pm=\pm B$ which are independent of $\gamma$. This leads to the following inference relating to the present situation.

\begin{inference}
1. If $\vert\psi_{\pm}^\gamma\rangle$ is an eigen vector of $H_\gamma$, $\mathcal{T}\vert\psi_\pm^{\gamma}\rangle$ is an eigenvector of $H^\dagger_\gamma$ and they are mutually orthogonal or in other words, \textbf{time reversed state vector is proportional to the bi-orthogonal partner of the state vector itself.} Furthermore, $\mathcal{T}\vert\psi_\pm^\gamma\rangle=(-1)^n\vert\psi_\mp^{-\gamma}\rangle$ with $n=0\:\:{\rm{or}}\:\:1$ depending on the choice of eigenvectors.

2. For $n=0$, $\vert\psi_{\pm}^\gamma\rangle$ and $\mathcal{T}\vert\psi_{\mp}^\gamma\rangle$ are eigenvectors of $H_\gamma$ and $H^\dagger_\gamma$ corresponding to  eigenvalues which are same in magnitude but opposite in sign. For $n=1$, $\vert\psi_{\pm}^\gamma\rangle$ and $\mathcal{T}\vert\psi_{\mp}^\gamma\rangle$ are eigenvectors of $H_\gamma$ and $H^\dagger_\gamma$ corresponding to  eigenvalues which are same in magnitude and  sign.	
\end{inference}

\begin{remark}
It would be conjectural in sense to claim that the above inference holds good for any set of bi-orthogonal states corresponding to a pseudo-Hermitian Hamiltonian that is isospectral to its Hermitian conjugate (both admitting  non-degenerate spectrum) and both are acting on a Hilbert space of  arbitrary dimension.
\end{remark}

\subsection{\textbf{Time reversal and Clifford algebra}}

\subsubsection{\textbf{Involutions and spinors as left ideal of $Cl_3$}}

\begin{dfn}
	A unary operation $\sharp$ acting on an element $u$ is said to be an involution iff $(u^{\sharp})^{\sharp}=u$. An involution may be an automorphism or an anti-automorphism according to the condition $(uv)^{\sharp}=u^{\sharp}v^{\sharp}$ or $(uv)^{\sharp}=v^{\sharp}u^{\sharp}$ respectively.
\end{dfn}

It is interesting to note that any arbitrary element $u\in Cl_3$ comprising of a scalar $\langle u\rangle_0$, a vector $\langle u\rangle_1$, a bivector $\langle u\rangle_2$ and a trivector $\langle u\rangle_3$ can have following three types of involution.

\begin{eqnarray*}
	\widehat{u}&=&\langle u\rangle_0-\langle u\rangle_1+\langle u\rangle_2-\langle u\rangle_3:\:\:{\rm{grade}}\:\:{\rm{inversion}}\:\:{\rm{(automorphism)}}\nonumber\\
	\widetilde{u}&=&\langle u\rangle_0+\langle u\rangle_1-\langle u\rangle_2-\langle u\rangle_3 : \:\:{\rm{reversion}}\:\:{\rm{(anti-automorphism)}}\nonumber\\
	\overline{u}&=&\langle u\rangle_0-\langle u\rangle_1-\langle u\rangle_2+\langle u\rangle_3 : \:\:{\rm{Clifford}}\:\:{\rm{conjugation}}\:\:{\rm{(anti-automorphism)}}
\end{eqnarray*}

 We shall make use of these involutions to observe various inner-products on $\mathcal{S}$, a minimal ideal of $Cl_3$.

 \begin{prop}
  Considering $u=\left(\begin{array}{cc}
	a_{11} & a_{12}\\
	a_{21} & a_{22}
\end{array} \right)\in Cl_3$ in terms of $\mathcal{G}$ the above involutions are equivalent to the following expressions

\begin{eqnarray}
	\widehat{u}=\left(\begin{array}{cc}
		a_{22}^{\star} & -a_{21}^{\star}\\
		-a_{12}^{\star} & a_{11}^{\star}
	\end{array} \right),\:\:\widetilde{u}=\left(\begin{array}{cc}
		a_{11}^{\star} & a_{21}^{\star}\\
		a_{12}^{\star} & a_{22}^{\star}
	\end{array} \right)	\:\:{\rm{and}}\:\:\overline{u}=\left(\begin{array}{cc}
		a_{22} & -a_{12}\\
		-a_{21} & a_{11}
	\end{array} \right)	
\end{eqnarray}
 \end{prop}

\textbf{Proof}
Writing $u$ in $Cl_3(R)$ generators we get $u=u_0\mathbf{1}_2+u_1e_1+u_2e_2+u_3e_3+u_{12}e_{12}+u_{13}e_{13}+u_{23}e_{23}+u_{123}e_{123}$. This gives

 \begin{eqnarray}
 {u}=\left(\begin{array}{cc}
		u_0+u_3+iu_{12}+iu_{123} & u_1-iu_2-u_{13}+iu_{23}\\
		u_1+iu_2+u_{13}+iu_{23} & u_0-u_3-iu_{12}+iu_{123}
	\end{array} \right)=\left(\begin{array}{cc}
		a_{11} & a_{12}\\
		a_{21} & a_{22}
	\end{array} \right)		
 \end{eqnarray}
Now it is obvious to note that $\widehat{a}_{11}=u_0-u_3+iu_{12}-iu_{123}$  holds by definition of grade involution and it is equal to $a_{22}^\star$. Similarly,  all the other terms can be verified for $\widehat{u}$. The same process may be repeated for the other two involutions. $\square$.

  \vspace{.5cm}

Let us write the following matrices with the help of $\mathcal{G}_{\gamma}$ and $\breve{\mathcal{G}}_{\gamma}=\{{1}, \breve{e}^{\gamma}_1, \breve{e}^{\gamma}_2, \breve{e}^{\gamma}_3, \breve{e}^{\gamma}_{12}, \breve{e}^{\gamma}_{23}, \breve{e}^{\gamma}_{31}, \breve{e}^{\gamma}_{123} \}=\{\mathbf{1}_2, -\widetilde{\sigma}_1^{\gamma}, -\widetilde{\sigma}_2^{\gamma}, -\widetilde{\sigma}_3^{\gamma}, i\widetilde{\sigma}_3^{\gamma}, i\widetilde{\sigma}_1^{\gamma}, i\widetilde{\sigma}_2^{\gamma}, -i\mathbf{1}_2\}$ (where, the later may be called the set of \textbf{time reversed generators} defined by $\breve{g}= \mathcal{T}^{-1}g\mathcal{T}$ for any generator $g\in Cl_3$).

\begin{eqnarray}
	g_0&=&\frac{1}{2}\mathbf{1}_2+\frac{\omega}{4}(e_3^{\gamma}-\breve{e}_3^{\gamma})=\left(\begin{array}{cc}
		1 & 0\\
		0 & 0
	\end{array} \right)\nonumber\\
g_1&=&\frac{1}{2}e_2^{\gamma}+\frac{\omega}{4}(e_{23}^{\gamma}+\breve{e}_{23}^{\gamma})=\left(\begin{array}{cc}
	0 & 0\\
	i & 0
\end{array} \right)\nonumber\\
g_2&=&\frac{1}{2}e_{31}^{\gamma}-\frac{\omega}{4}(e_1^{\gamma}-\breve{e}_1^{\gamma})=\left(\begin{array}{cc}
	0 & 0\\
	-1 & 0
\end{array} \right)\nonumber\\
g_3&=&\frac{1}{2}e_{123}+\frac{\omega}{4}(e_{12}^{\gamma}+\breve{e}_{12}^{\gamma})=\left(\begin{array}{cc}
	i & 0\\
	0 & 0
\end{array} \right)
\end{eqnarray}

With $\{\xi_0, \xi_1, \xi_2, \xi_3\}\in\mathbf{R}$ we can write an element $\Psi=\sum_{j=0}^{3}\xi_jg_j$ of the left ideal  $\mathcal{S}$ (of $Cl_3$) in terms of the generators $\mathcal{G}_s=\{g_j\vert j=0, 1, 2, 3\}$ derived from bi-orthogonal system. Identifying $\psi_1=\xi_0+i\xi_3$ and $\psi_2=-\xi_2+i\xi_1$ a typical spinor state ${\psi}=\left(\begin{array}{c}
	\psi_1 \\
	\psi_2
\end{array} \right)\in\mathbf{C}^2$ can be cast into $2\times 2$ matrix form ${\Psi}=\left(\begin{array}{cc}
\psi_1 & 0\\
\psi_2 & 0
\end{array} \right)\in M(2, \mathbf{C})g_0$, where, $M(2, \mathbf{C})$ is the space of $2\times 2$ matrices over $\mathbf{C}$, $g_0=\left(\begin{array}{cc}
1 & 0\\
0 & 0
\end{array} \right)$ is an idempotent $g_0^2=g_0$ and $M(2, \mathbf{C})g_0\cong \mathbf{C^2}$. $\Psi\in\mathcal{S}=M(2, \mathbf{C})g_0\subset {Cl}_3$ is the minimal left ideal of ${Cl}_3$ in the sense that any $u\in {Cl}_3$ and $\Psi\in\mathcal{S}$, $u\Psi\in \mathcal{S}$.

The effect of time reversal in $\mathcal{S}$ itself can also be understood in terms of a unary operation $'\flat'$ which is equivalent to \textbf{basis flip} as given below
\begin{equation}
\left.\begin{array}{c}
\mathbf{1}_2\rightarrow e_{13}\\
e_{13}\rightarrow -\mathbf{1}_2
\end{array} \right\}, \left.\begin{array}{c}
e_1\rightarrow -e_{3}\\
e_{3}\rightarrow e_1
\end{array} \right\}, \left.\begin{array}{c}
e_2\rightarrow e_{123}\\
e_{123}\rightarrow -e_2
\end{array} \right\}\:\:{\rm{and}}\:\:\left.\begin{array}{c}
e_{12}\rightarrow -e_{23}\\
e_{23}\rightarrow e_{12}
\end{array} \right\}.
\end{equation}

 This makes $u^\flat=u_0e_{13}-u_1e_3+u_2e_{123}+u_3e_1-u_{12}e_{23}+u_{23}e_{12}-u_{13}\mathbf{1}_2-u_{123}e_2=\left(\begin{array}{cc}
-a^\star_{21} & -a^\star_{22}\\
a^\star_{11} & a^\star_{12}
\end{array} \right)$ and $(u^\flat)^\flat=-u$. This makes $\Psi^{\flat}$ the time reversed version of $\Psi$ and also an element of the left ideal $\mathcal{S}$. It can be verified that $(\Psi^{\flat})^{\flat}=-\Psi$ and this fact is consistent with the anti-involution property of fermionic time reversal .

\begin{remark}\label{invo}
The projection operators that has been studied in section-\ref{bi} are related by grade involution. It can be readily verified that $\widehat{\Pi}_\pm^\gamma=\Pi_\mp^{-\gamma}$.  This means while the time reversed Hamiltonian is nothing but the \textbf{reversion} of the actual Hamiltonian in the present case it may also be possible to develop a parallel quantum mechanics in terms of \textbf{ grade involuted projectors} which are essential for measurement. Construction of a bi-orthogonal quantum mechanics and the interpretation of measurement have already been elaborately discussed in \cite{brody14}. In the present scenario it is suggestive to define the so called \textbf{associated  state} $\vert\widecheck{\psi}\rangle={\alpha}^{-\gamma}_+\vert\psi_+^{-\gamma}\rangle+{\alpha}^{-\gamma}_-\vert\psi_-^{-\gamma}\rangle$ corresponding to a given set of states $\vert\psi\rangle={\alpha}^{\gamma}_+\vert\psi_+^{\gamma}\rangle+{\alpha}^{\gamma}_-\vert\psi_-^{\gamma}\rangle$ and an inner-product $\langle{\phi}, \psi\rangle=\langle \widecheck{\phi}\vert\psi\rangle=(\beta^{-\gamma}_+)^\star {\alpha}^{-\gamma}_++(\beta^{\gamma}_+)^\star {\alpha}^{-\gamma}_-$. The expectation value of any observable $F$ in a state $\vert\psi\rangle$ is given by $\langle F\rangle=\frac{\langle\widecheck{\psi}\vert F\vert\psi\rangle}{\langle\widecheck\psi\vert\psi\rangle}$.
\end{remark}

\subsubsection{\textbf{Involutions and inner-products}}
\par The bi-orthogonality that has been observed in section-\ref{bi} for $\Psi,\Phi\in\mathcal{S}$ can now be understood in two ways.

\textbullet\:\:\textbf{Reversion}

Let us consider the matrix product
 \begin{eqnarray}
 	(\widetilde{\Psi}\Phi)=\left(\begin{array}{cc}
 		\psi_{1}^{\star} & \psi_{2}^{\star}\\
 		0 & 0
 	\end{array} \right)	\left(\begin{array}{cc}
 	\phi_{1} & 0\\
 	\phi_{2} & 0
 \end{array} \right)=\left(\begin{array}{cc}
 \psi_{1}^{\star}\phi_{1}+\psi_{2}^{\star}\phi_{2} & 0  \\
 0 & 0
\end{array} \right)	
 \end{eqnarray}
 $(\widetilde{\Psi}\Phi)=\left(\begin{array}{cc}
	0 & 0  \\
	0 & 0
\end{array} \right)$ implies the said bi-orthogonality. If the  representations of $\Psi$ and $\Phi$ are replaced by their respective time-reversed versions $\Psi^\flat$ and   $\Phi^\flat$ it is easy to verify that $(\widetilde{\Phi}^{\flat}\Psi^{\flat})$ leads to the same conclusion.

\begin{remark}
The inner-product defines a map $\mathcal{S}\times \mathcal{S}\rightarrow \mathcal{R}$ where, $\mathcal{R}=g_0Cl_3g_0=\left\{\left(\begin{array}{cc}
	c & 0  \\
	0 & 0
\end{array} \right) : c\in\mathbf{C}\right\}$ is a division ring with $g_0\in Cl_3$ being the primitive. It is also a division algebra isomorphic to $\mathbf{C}$.	
\end{remark}
\begin{remark}
	The right $\mathcal{R}$-linear transformations $\psi\rightarrow s\psi$, $\:\:\forall\:\:s\in Cl_3$ being a spinor constitutes an automorphism group of the scalar product which is preserved under its action that is $(\widetilde{s\psi})s\phi=\widetilde{\psi}\phi$ and furthermore $\{s\in Cl_3 : \tilde{s}s=\mathbf{1}_2\}\cong U(2)=\{s\in M(2, \mathbf{C}) : s^{\dagger}s=\mathbf{1}_2\}$.
\end{remark}

\textbullet\:\:\textbf{Clifford Conjugation coupled with grade inversion}

The conventional inner product can also be expressed involving time reversal into account and that reflects into the process of Clifford involution. Let us write $\langle\langle\phi\vert\psi\rangle\rangle=(-\phi_2\:\:\phi_1)\mathcal{T}\left(\begin{array}{c}
		\psi_1  \\
		\psi_2
	\end{array} \right) =\phi_1\psi_1^\star+\phi_2\psi_2^\star$. Such an inner product justifies anti-uniterity of $\mathcal{T}$ i. e.; $\langle\langle\mathcal{T}\psi\vert\mathcal{T}\phi\rangle\rangle=\langle\langle\phi\vert\psi\rangle\rangle$. Right-multiplying Clifford conjugation of $\Phi$ with grade inversion of $\Psi$ one can write
\begin{eqnarray}\label{cliffconju}
	\overline{(\overline{\Phi}\widehat{\Psi})}=\overline{\left(\begin{array}{cc}
		0 & 0  \\
		-\phi_2 & \phi_1
	\end{array} \right) \left(\begin{array}{cc}
	0 & -\psi^\star_2  \\
	0 & \psi^\star_1
\end{array} \right)}=\left(\begin{array}{cc}
\phi_1\psi_1^\star+\phi_2\psi_2^\star & 0  \\
0 & 0
\end{array} \right)
\end{eqnarray}.
 The final element in equation-\ref{cliffconju} very much belongs to the ring $\mathcal{R}$.

\begin{remark}
 It can be verified that $(\widetilde{\Psi}_+^{\gamma}\Psi_-^{-\gamma})=0=\overline{(\overline{{\Psi}_+^{\gamma}}\widehat{\Psi}_-^{-\gamma})}$ and $(\widetilde{\Psi}_-^{\gamma}\Psi_+^{-\gamma})=0=\overline{(\overline{{\Psi}_-^{\gamma}}\widehat{\Psi}_+^{-\gamma})}$.
\end{remark}

\subsubsection{\textbf{Spinor operator and Time reversal}}

Time reversal operator $\mathcal{T}$ has more involved physical meaning when the Spinor operator is understood as the element of the even subalgebra $Cl_3^+$ in terms of the present generators $\mathcal{G}_+^{\gamma}=\{1, e_{12}^{\gamma}, e_{23}^{\gamma}, e_{31}^{\gamma}\}$ and the corresponding time reversed generators
$\breve{\mathcal{G}}_+^{\gamma}=\{1, \breve{e}_{12}^{\gamma}, \breve{e}_{23}^{\gamma}, \breve{e}_{31}^{\gamma}\}$. Considering a set  $\{\Xi_0, \Xi_{12}, \Xi_{23}, \Xi_{31}\}\in\mathbf{R}$, let us write $\mathbf{\Xi}_{\gamma}$  and its time reversed version $\breve{\mathbf{\Xi}}_{\gamma}$ as
\begin{equation*}
	\mathbf{\Xi}_{\gamma}=\Xi_{0}\mathbf{1}+\omega\Xi_{12}e_{12}^{\gamma}+\omega\Xi_{23}e_{23}^{\gamma}+\Xi_{31}e_{31}^{\gamma}\:\:{\rm{and}}\:\:
	\breve{\mathbf{\Xi}}_{\gamma}=\Xi_{0}\mathbf{1}+\omega\Xi_{12}\breve{e}_{12}^{\gamma}+\omega\Xi_{23}\breve{e}_{23}^{\gamma}+\Xi_{31}\breve{e}_{31}^{\gamma}\:\:{\rm{respectively.}}
\end{equation*}
This gives
\begin{eqnarray}
	\mathbf{\Xi}&=&\frac{1}{2}(\mathbf{\Xi}_{\gamma}+\breve{\mathbf{\Xi}}_{\gamma})=\left(\begin{array}{cc}
		\Xi_0+i\Xi_{12} & \Xi_{31}+i\Xi_{23} \\
		-\Xi_{31}+i\Xi_{23} & \Xi_0-i\Xi_{12}
	\end{array} \right)\nonumber\\
&=&\left(\begin{array}{cc}
	\Xi_1 & -\Xi_2^{\star} \\
	\Xi_2 & \Xi_1^{\star}
\end{array} \right)=\left(\begin{array}{cc}
\Xi_1 & 0 \\
\Xi_2 & 0
\end{array} \right)+\left(\begin{array}{cc}
0 & -\Xi_2^{\star} \\
0 & \Xi_1^{\star}
\end{array} \right)=\Xi+\widehat{\Xi}.
\end{eqnarray}
It is easy to observe that the non-zero columns of $\widehat{\Xi}$ (the grade involution of $\Xi$) is the time-reversed expression of the non-zero column of $\Xi$. $\mathbf{\Xi}$ is called the Spinor Operator \cite{Lo01, vaz19}. The expression $\Upsilon_-=\frac{\omega}{2}\mathbf{\Xi}(e_3^\gamma-\breve{e}_3^{\gamma})\widetilde{\mathbf{\Xi}}=\mathbf{\Xi} e_3\widetilde{\mathbf{\Xi}}$ gives three distinct real numbers leading to \textbf{Kustaanheimo-Stiefel (KS) transformation} \cite{Lo01, cahill90, chen00, helff96, ino93}.
Taking  $\Xi_1=r_1+ir_2$ and $\Xi_2=r_3+ir_4$, $r_1, r_2\in\mathbf{R}$, this transformation $\{x_1, x_2, x_3\}\rightarrow\{r_1, r_2, r_3, r_4\}$ ($\mathbf{R}^3\rightarrow\mathbf{R}^4$) implies
\begin{eqnarray}
x_1=2(r_2r_3-r_1r_4),\:\: x_2=2(r_1r_3+r_2r_4)\:\:{\rm{and}}\:\: x_3=r_1^2+r_2^2-r_3^2-r_4^2.
\end{eqnarray}
 In order to set an equivalence with spherical polar coordinates $\{x_1=r\sin\theta\cos\phi, x_2=r\sin\theta\sin\phi, x_3=r\cos\theta\}$ the possible choice of $\{r_1, r_2, r_3, r_4\}$ can be given by

 \begin{equation}
 \left.\begin{array}{c}
r_1=\sqrt{r}\cos\frac{\theta}{2}\sin\omega_1\\
r_3=\sqrt{r}\sin\frac{\theta}{2}\cos\omega_2
\end{array} \right. \:\:\left.\begin{array}{c}
r_2=\sqrt{r}\cos\frac{\theta}{2}\cos\omega_1\\
r_4=\sqrt{r}\sin\frac{\theta}{2}\sin\omega_2
\end{array} \right.\:\:{\rm{with}}\:\:\phi=\omega_1+\omega_2.
 \end{equation}

KS transformation has been used to understand three dimensional Kepler problem in terms of four dimensional harmonic oscillator \cite{chen00}. However similar realization is possible from the expression $\Upsilon_+=\frac{\omega}{2}\mathbf{\Xi}(e_3^\gamma+ \breve{e}_3^{\gamma})\widetilde{\mathbf{\Xi}}$ that implies the following choice

\begin{eqnarray}
x_1=2(r_1r_2+r_3r_4),\:\: x_2=2(r_1r_3-r_2r_4)\:\:{\rm{and}}\:\: x_3=r_1^2-r_2^2-r_3^2+r_4^2.
\end{eqnarray}

Choosing
\begin{equation}
 \left.\begin{array}{c}
r_1=\sqrt{r}\cos\frac{\theta}{2}\sin\omega_1\\
r_3=\sqrt{r}\sin\frac{\theta}{2}\cos\omega_2
\end{array} \right. \:\:\left.\begin{array}{c}
r_2=\sqrt{r}\sin\frac{\theta}{2}\sin\omega_2\\
r_4=\sqrt{r}\cos\frac{\theta}{2}\cos\omega_1
\end{array} \right.\:\:{\rm{with}}\:\:\phi=\omega_1-\omega_2.
 \end{equation}
we get another possibility of KS-transformation.

\section{Comment}
According to Brody \cite{brody14}, apart from the requirement of completeness (which is implied in finite dimension) and some mathematical conveniences, there is hardly any compulsion to use Hermitian operator and allied orthogonality of eigenvectors to justify a possible quantum formalism. Bi-orthogonal quantum mechanics stands in favour of this idea. However, there are issues like space-time reflection symmetry, Kramers theorem, situations involving infinite dimension (a free particle case is under preparation) and many other which are yet to be adequately addressed in such a frame work. A Kramers like phenomenon, however conjectural in sense, has been envisaged in section-\ref{pseudo} and the role of time-reversal to connect an eigen-spinor and its bi-orthogonal partner has been made visible. It has been observed that the condition that restricts the possibility of a typical one parameter bi-orthogonality to construct Clifford generators remains to be relevant even to the construction of two-parameter bi-orthogonal eigen-spinors of the Hamiltonians discussed.

Secondly, much stress has been given on the connections of fermionic time reversal  and various definitions of inner-product in the light of involutions admitted by Clifford algebra. The very attempt of viewing spinor as left ideal of Clifford algebra and the concept of time reversed generator introduced here seems to have enough potential to explore the role of geometric algebra in the recent surge of pseudo-Hermitian quantum mechanics. The newly introduced variant of KS-transformation is indicative of the richness of pseudo-Hermitian methods in some applicational aspects of  Clifford algebra.   The possibility of a quaternionic interpretation \cite{finkel62, adler86} of the similar scenario even in the frame work of real Hilbert space\cite{giardino20} may not be ruled out.

\vspace{.5cm}

\noindent {\bf \Large Acknowledgement}

A.C. wishes to thank his colleague Dr. Baisakhi Mal for her valuable assistance in preparing the latex version.

\vspace{.5cm}

\noindent {\bf \Large Data Availability Statement} : Data sharing not applicable to this article as no datasets were generated or analysed during the current study.

\vspace{.5cm}

\noindent {\bf \Large Conflict of interest statement} : 	The authors have no relevant financial or non-financial interests to disclose.

\end{document}